\documentclass[aps,prl,twocolumn,superscriptaddress,showpacs]{revtex4}
\usepackage{graphics,graphicx}
\usepackage{color}
\usepackage{amssymb,bm} \usepackage{amsmath}
\usepackage{bookmath}
\begin{document}
%~~~~~~~~~~~~~~~~~~~~~~~~~~~~~~~~~~~~~~~~~~~~~~~~~
\title{Surface bound states and spin currents
in non-centrosymmetric superconductors}
\author{A. B. Vorontsov}\altaffiliation{
\vspace{-0.5cm}
Present address: Dept. of Physics, University of Wisconsin, Madison, WI 53706}
\author{I. Vekhter}
\affiliation{Department of Physics and Astronomy,
             Louisiana State University, Baton Rouge, Louisiana, 70803, USA}
\author{M. Eschrig}
\affiliation{Institut f\"ur Theoretische Festk\"orperphysik and
DFG-Center for Functional Nanostructures, Universit\"at Karlsruhe,
D-76128 Karlsruhe, Germany}
\date{\today}
%\keywords{}

\begin{abstract}
We investigate the influence of spin-orbit coupling in a
non-centrosymmetric superconductor on its ground state properties
near a surface. We determine the spectrum of Andreev bound states
due to surface-induced mixing of bands with opposite spin
helicities for a Rashba-type spin-orbit coupling. We find
a qualitative change of the Andreev spectrum when we account for the
suppression of the order parameter near the surface, leading to
clear signatures in the surface density of states. We also
compute the spin current at the surface, which has spin
polarization normal to that of the bulk current. The magnitude of
the current at the surface is enhanced in the normal state
compared to the bulk, and even further enlarged in the
superconducting phase. The particle and hole coherence amplitudes
show Faraday-like rotations of the spin along quasiparticle
trajectories.
\end{abstract}
\maketitle
%~~~~~~~~~~~~~~~~~~~~~~~~~~~~~~~~~~~~~~~~~~~~~~~~~~~~~~~~~~~~~~~~~~~~~~~~~~~~~
%~~~~~~~~~~~~~~~~~~~~~~~~~~~~~~~~~~~~~~~~~~~~~~~~~~~~~~~~~~~~~~~~~~~~~~~~~~~~~

%\paragraph {\it Introduction.}

The role of chirality and spin-orbit coupling in materials and
nanostructures is a very active and promising subject in the
fields of spintronics, superconductivity and magnetism
\cite{bode07,meier07,champel08}. The unusual properties of
non-centro\-symmetric (NCS) materials originate from the crystal
structure that lacks a center of inversion, allowing for
pronounced spin-orbit (SO) coupling that is odd in the
electron momentum, and leading to a chiral ground state.
The resulting two-band nature of NCS metals
leads to effects reminiscent of semiconductor physics, such as
birefringence and spin polarization of the electron
wavepacket \cite{kho04}.
Especially promising is the presence of charge-neutral spin
currents
in the ground state~\cite{SMurakami:2003,EIRashba:2003}.

Since understanding of interface physics is one of the foundations
for all potential applications, it is of pivotal interest to
investigate how the physical properties of NCS materials are
modified near surfaces. The key observation is that scattering
events off interfaces in materials with strong spin-orbit effects
are typically spin-active. Spins dominate the surface physics,
and any successful theoretical treatment must take this into
account.

The recently discovered class of NCS superconductors
\cite{bau04,yua06,TAkazawa:2004,KSugawara:2007} combines the
strong SO coupling that governs the metallic bands with a
non-trivial, chiral, spin structure of the superconducting (SC) order
parameter \cite{gor01,fri04,fri06}. As a
result, one may expect that spin transport in the SC phase
exhibits novel features compared to superconductors with negligible
SO interaction. These
features are expected to be especially prominent near surfaces and interfaces,
where the physics is controlled by the Andreev bound states (ABS),
built as a result of particle-hole coherent scattering.  ABS are
crucially important in unconventional
superconductors~\cite{lof01}, where the phase variation of the
order parameter (OP) on the Fermi surface \cite{sig91} and the
pairbreaking near interfaces may lead to a midgap peak in the density
of states (DOS) at the surface. The ABS states control
thermodynamic properties and stability of the surface phases
\cite{mat95,buc95a}, and govern transport across interfaces
\cite{cov97,fog97a,apr98}.

In this Letter we study the Andreev states and spin currents
at the surface of a NCS superconductor. We show that (i) the ABS
spectrum is qualitatively modified by the self consistent
suppression of the order parameter; (ii) the spin current is
(a) strongly enhanced near the surface in the normal state; (b)
further enhanced in the SC phase. We develop a detailed theory of
these effects.

\begin{figure}[t]
\begin{center}
\includegraphics[width=7cm]{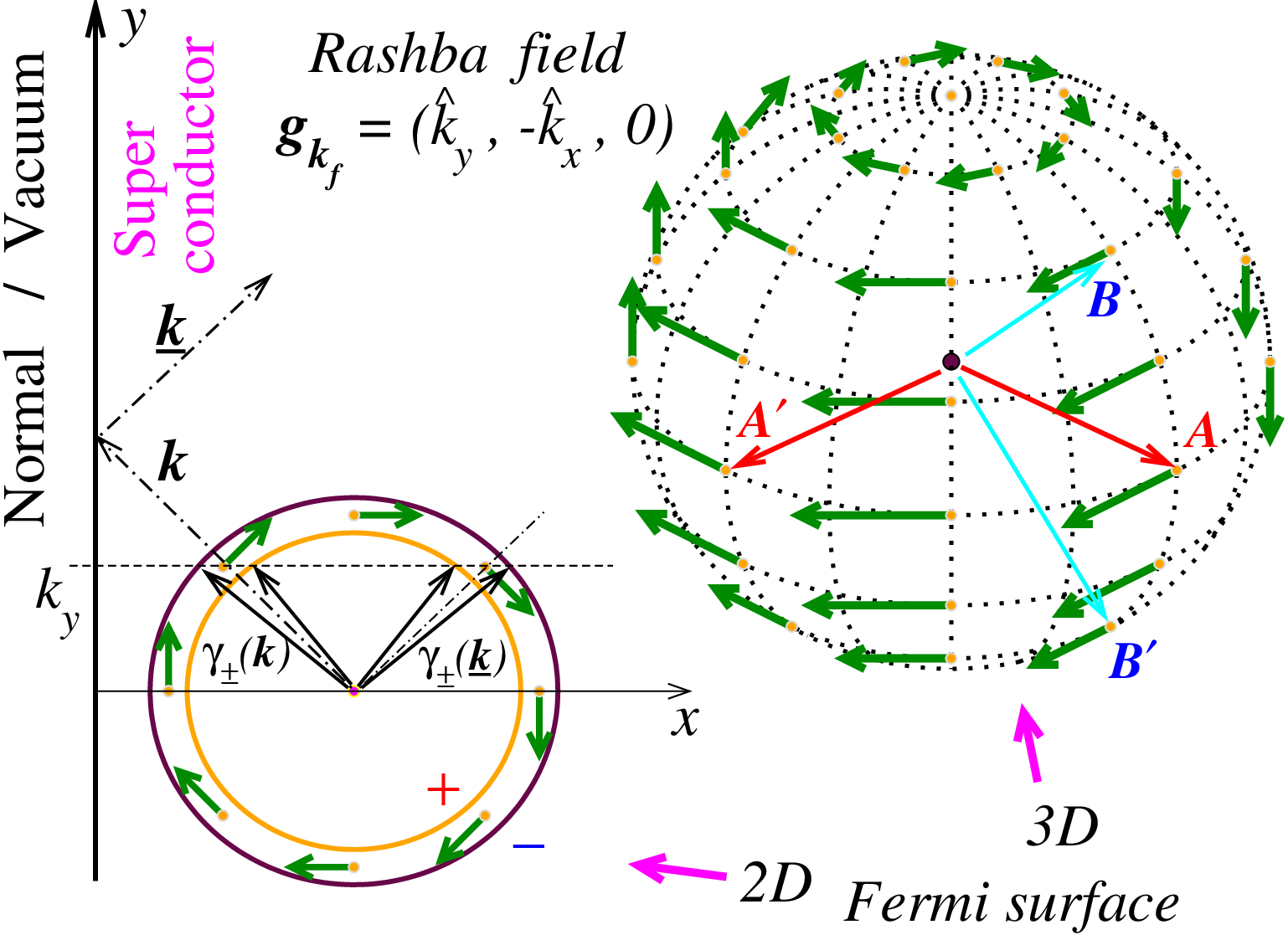}
\end{center}
\caption{\label{fig:1} (Color online) A map of the spin-orbit
vector in momentum space for the Rashba form
$\vg_\vk=\hat\vk\times \hat\vz$.
On reflection the spin orbit vector
$\vg_\vk$ may change, e.g. from $A \to A'$, or not, $B\to B'$.
The scattering geometry is shown on the left. }
\end{figure}

For a non-centrosymmetric material it is convenient to perform a
canonical transformation from a spin basis (with fermion
annihilation operators $c_{\vk \mu }$ for spin
$\mu=\uparrow,\downarrow $) to the so-called helicity basis
($b_{\vk s}$ with helicity $s=\pm $), that diagonalizes the
kinetic part of the Hamiltonian,
    \be
\cH_{kin} = \sum_{\vk \mu \nu } c^\dag_{\vk \mu }
(\xi^{\; }_\vk + \alpha \vg^{\; }_\vk \vsigma)_{\mu \nu} c^{\; }_{\vk \nu}
 = \sum_{\vk s} \vare^{\; }_{\vk s}  b^\dag_{\vk s} b^{\; }_{\vk s} \,.
\label{eq:HN}
    \ee
Here, $\xi_\vk $ is the band dispersion
relative to the chemical potential in the absence of SO
interaction, $\alpha $ is the SO coupling strength, $\vsigma $
is the vector of Pauli matrices, and $\vg_\vk $ is a normalized
(see below)
SO vector \cite{gor01,fri04} that is
odd in momentum, $\vg_{-\vk} =-\vg_\vk$, see Fig.~\ref{fig:1}. The
helicity band dispersion is $\vare_{\vk \pm} = \xi_\vk \pm \alpha
|\vg_\vk| $. SO interaction fixes the orientation of the
quasiparticle spin with respect to its momentum in each helicity
band.

%%%%%%%%%%%%%%%%%%%%%%%%%%%%%%%%%%%%%%%%%%%%%%%%%%%%%%%%%%%%%
%%%%%%%%%%%%%%%%%%%%%%%%%%%%%%%%%%%%%%%%%%%%%%%%%%%%%%%%%%%%%

The Hamiltonian \eqref{eq:HN} is time reversal invariant but
lifts the spin degeneracy. The
transformation from spin to helicity basis, $U_\vk $, is defined
by $U^{\; }_\vk (\vg^{\; }_\vk \vsigma)U^\dagger_\vk = |\vg_\vk
|\sigma_3$, and determined by the direction of the $\vg$-vector in
$\vk$ space,
    \be U_\vk = e^{-i
\frac{\theta_\vg }{2} \displaystyle\vn_\vg \vsigma }, \quad
\vn_\vg = \frac{\vg_\vk \times \hat\vz }{|\vg_\vk \times \hat\vz
|}\,,
    \label{eq:U}
    \ee
where $\hat\vz $ is the unit vector in z-direction, and
$\theta_\vg$ is the polar angle between $\vg_\vk$ and $\hat\vz $
\cite{fri04}.

To describe superconductivity we use the Nambu-Gor'kov formalism
modified for a helical basis. We define the helical counterpart,
$\hat B^\dag_{\vk } = (b^\dag_{\vk +},
b^\dag_{\vk -}, b^{\; }_{\vk +},
b^{\; }_{\vk -})$,
to the Nambu spinor, $\hat C^\dagger_{\vk } = (c^\dagger_{\vk \uparrow},
c^\dagger_{\vk \downarrow}, c^{\; }_{\vk \uparrow},
c^{\; }_{\vk \downarrow})$,
by
\be
\hat B_{\vk } = \hat U_\vk \hat C_{\vk }, \qquad \hat{U}  =  \left(
\begin{array}{cc} U & 0 \\ 0 & U^* \end{array} \right) \,
    \ee
and construct $4\times 4$ retarded Green's functions in helicity basis,
$
\hat G_{\vk_1 \vk_2}(t_1,t_2)=
-i\theta(t_1-t_2) \langle
\big\{ \hat B_{\vk_1}(t_1), \hat B^\dagger_{\vk_2} (t_2) \big\} \rangle_{\cal H}
$,
where $\hat B(t)$
are Heisenberg operators,
the braces denote an anticommutator,
$\langle \ldots \rangle_{\cal H}$ is a grand canonical average,
and $\theta $ is the usual step function.

Below we employ the quasiclassical method \cite{JWSerene:1983} for
treating the inhomogeneous surface problem. In the materials of
interest $\alpha |\vg_{\vk_f}| \ll E_f $ for any Fermi momentum
$\vk_f$, where $E_f$ is the Fermi energy. In addition, the
superconducting energy scales (transition temperature $T_c$ and
the gap $\Delta $) are much smaller than $E_f$. Under these
conditions quasiparticles with different helicity but with the
same $\hvk \equiv \vk /|\vk |$ propagate 
coherently along a common classical trajectory 
(determined by the Fermi surface for $\alpha=0$, $\xi_{\vk_f}=0$), 
over distances much longer than the Fermi wavelength.
%, and can be assigned to a common Fermi surface.
We normalize $\vg_\vk $, $\langle\vg^2_{\vk_f}\rangle=1$, where
$\langle \ldots \rangle$ denotes a Fermi surface average. The
quasiclassical propagator
is then obtained as
$\hat g(\vk_f,\vR, \epsilon,t)= \hat \tau_3 \int d\xi_\vk \int
(d\vq )(d\tau ) e^{i(\vq \vR+\epsilon \tau )} \hat
G_{\vk+\frac{\vq}{2},\vk-\frac{\vq}{2}} (t+\frac{\tau
}{2},t-\frac{\tau }{2}) $ where $\hat\tau_3$ is the Pauli matrix
in the particle-hole space.
Using $U_{-\vk} U_\vk^\dag = i \vn_\vg \vsigma $
and the fermionic anticommutation relations for the $b$ and $b^\dagger$,
we derive the fundamental symmetry relations for the
2x2 Nambu matrix components,
$g(\epsilon, \vk_f )_{22} = \left[ (\vn_\vg \vsigma ) \;
g(-\epsilon, -\vk_f )_{11} \; (\vn_\vg \vsigma ) \right]^\ast$ and $
g(\epsilon, \vk_f )_{21} = \left[ (\vn_\vg \vsigma ) \; g(-\epsilon,
-\vk_f )_{12} \; (-i \sigma_2) (\vn_\vg \vsigma) \right]^\ast
(i\sigma_2) $.

Standard procedure \cite{JWSerene:1983} yields the
Eilenberger equation in helicity basis,
    \be [\vare \hat{\tau}_3 -\alpha \hat v_{SO} -
\hDelta \;,\; \hg] + i\vv_f \grad \hg = \hat 0 \label{eq:eil}
    \ee
with normalization $\hg^2 = - \pi^2 \hat{1}$. Here, $\vare$ is the
energy, $\hat v_{SO}= |\vg_{\vk_f }|\, \sigma_3\hat\tau_3 $, and
$\hat \Delta $ is the
superconducting OP.
The velocity renormalization of order $\alpha /E_f\ll 1$ is neglected.
We choose a separable pairing interaction consistent
with the form of the gap, and determine $\hDelta$ self
consistently with $\hg$.
In NCS superconductors the OP is a mixture of spin singlet
($\Delta_s$) and triplet ($\Delta_t$) components
\cite{ser04,fri06}. 
Assuming that the triplet component aligns with $\vg_\vk $,
in real gauge it is given by, 
\be \hDelta =
\cY(\vk_f ) \left[ \Delta_s( \vR )  \hat 1 + \Delta_t(\vR ) \hat
v_{SO}(\vk_f)\right] \; (i\sigma_2) \hat\tau_1 \,, \label{eq:OP}
\ee
where the basis function $\cY(\vk_f)$
transforms according to one of the irreducible representations of
the crystal point group, and $\langle\cY^2(\vk_f)\rangle=1$. With
the gap functions in the helicity bands,
$\Delta_\pm=\Delta_s\pm\Delta_t |\vg_{\vk_f}|$, the order
parameter is $\Delta = \left\{\Delta_+ \sigma^+ -\Delta_- \sigma^-
\right\}\cY$, where $\sigma^\pm=(\sigma_1\pm i\sigma_2)/2$.

We parameterize the Green's function by the coherence functions
for particles and holes, $\gamma $ and $\bar\gamma$ (2x2 spin
matrices), which allow a very intuitive physical interpretation of
the Andreev scattering processes \cite{esc00},
    \be
\hspace*{-0.25cm}
    \hg= -i\pi
\left(\begin{array}{cc} 1-\gamma {\bar\gamma} & 0\\ 0&
1-{\bar\gamma} \gamma
\end{array} \right)^{-1} \left(\begin{array}{cc}
1+\gamma {\bar\gamma} & 2 \gamma  \\ -2 {\bar \gamma}& -1+{\bar
\gamma} \gamma \end{array}\right). \label{eq:sgf}
    \ee
Fundamental symmetry relates $\gamma $ and $\bar\gamma$ in helicity basis by
$\bar\gamma (\epsilon,\vk_f )= \left[ (\vn_\vg \vsigma )
\gamma (-\epsilon,-\vk_f ) (-i\sigma_2 ) (\vn_\vg \vsigma
)\right]^\ast (i\sigma_2)$.
In the bulk,
$\gamma=\gamma^\sm{0}_+\sigma^+ - \gamma^\sm{0}_- \sigma^-$,
and $\bar\gamma=\tilde\gamma^\sm{0}_- \sigma^+ -\tilde\gamma^\sm{0}_+\sigma^-$,
with
$\gamma^\sm{0}_\pm(\epsilon,\vk_f)=-\Delta_\pm(\vk_f)/
(\epsilon+i\sqrt{|\Delta_\pm(\vk_f)|^2-\epsilon^2})$ and
$\tilde\gamma^\sm{0}_\pm(\vk_f,\epsilon)=
\gamma^\sm{0}_\pm (-\vk_f,-\epsilon)^\ast$.

The surface bound states are determined by the poles of the
Green's function, Eq.~\eqref{eq:sgf}. We consider specular
reflection, whereby the component of $\vk$ normal to surface
changes sign, $\vk\to\underline\vk$, see Fig.~\ref{fig:1}. We
find the amplitudes $\gamma_\vk$ ($\bar\gamma_{\ul\vk}$),
by integrating forward (backward) along incoming, $\vk$,
(outgoing, $\ul\vk$) trajectory starting from the values in the
bulk \cite{esc00}.
The amplitudes $\gamma_{\ul\vk}$ and
$\bar\gamma_\vk$, in contrast, are determined from the
boundary conditions at the surface.
Since the surface is non-magnetic, the
components of $\hg$ in the spin basis,
$\hg^\sm{s}_\vk =\hat{U}^\dagger_{\vk} \hg^{\; }_\vk \hat{U}^{\; }_{\vk}$,
are continuous at the surface.
This leads to a surface induced
mixing of the helicity bands according to $U^\dag_{\underline\vk}
\gamma^{\; }_{\underline\vk} U^\ast_{\underline\vk} =
\gamma^\sm{s}_{\underline\vk} = \gamma^\sm{s}_{\vk}= U^\dag_{\vk}
\gamma^{\; }_{\vk} U^\ast_{\vk}$, and $U^T_{\vk}
{\bar\gamma}^{\; }_{\vk} U^{\; }_{\vk}= {\bar\gamma}^\sm{s}_{\vk}=
{\bar\gamma}^\sm{s}_{\underline\vk}= U^T_{\underline\vk}
{\bar\gamma}^{\; }_{\underline\vk} U^{\; }_{\underline\vk}$. From
Eq.~(\ref{eq:sgf}), the bound states correspond to the zero
eigenvalues of the matrix
$ 1 - \gamma_\vk \bar\gamma_\vk= 1 -
\gamma^{\; }_\vk
U^\ast_{\vk } U^T_{\underline\vk}
\bar\gamma^{\; }_{\ul\vk}
U^{\; }_{\underline\vk} U^\dag_{\vk} $
at the surface,
and we derive our final equation for the ABS energies via
the surface amplitudes in the helicity basis \be (1 +
\gamma_+\tilde\gamma_+)(1 + \gamma_- \tilde\gamma_-) = -(1+
\gamma_+\tilde\gamma_-)(1 + \gamma_-\tilde\gamma_+) \cM \,.
\label{eq:BS} \ee The ``mixing'' factor $\cM$ is determined by the
change of $\vg_\vk $ under reflection $\vk \to \underline\vk $ at
the surface,
 \be
 \cM =
\frac{ \sin^2{\theta_\vg-\theta_{\ul\vg} \over 2} +
\sin^2{\theta_\vg+\theta_{\ul\vg} \over 2} \tan^2{\phi_\vg -\phi_{\ul\vg} \over 2} }
{ \cos^2{\theta_\vg-\theta_{\ul \vg} \over 2} +
\cos^2{\theta_\vg+\theta_{\ul\vg} \over 2} \tan^2{\phi_\vg -\phi_{\ul\vg} \over 2} } \,,
    \ee
where $\theta_\vg,\phi_\vg $ and $\theta_{\ul\vg},\phi_{\ul\vg}$
are the polar and azimuthal angles of $\vg_\vk $ and $\vg_{\ul\vk}$,
respectively.
If $\vg_{\vk }=\vg_{\underline\vk }$  ($B\to B'$ in Fig.~\ref{fig:1})
there is no helicity band mixing, $\cM=0$, and we recover the
conditions for ABS in superconductors with no SO coupling.
The limit $\cM\to \infty$ describes pure interband
scattering. In the general case ($A \to A'$ in Fig.~\ref{fig:1}) a
finite $\cM$ determines the relative weights of intraband and
interband scattering.

%\paragraph{Self-consistent determination of the order parameter.}
While assuming a uniform OP up to the surface to obtain the
ABS spectra may seem reasonable, we show now that the suppression
of the anisotropic (triplet) component of the OP in
Eq.~(\ref{eq:OP}) near the surface drastically modifies the ABS
spectrum and the surface DOS, $N(\vare,\vk_f)= -\frac{N_f}{2\pi}
\mbox{Im} \mbox{Tr}\left\{ g(\vare,\vk_f)\right\}$, where Tr is a
2x2 spin trace, and $N_f$ is the normal state DOS. Hereafter
we consider a 2D material with the Rashba type SO coupling $\alpha
= \alpha_R k_f,\,\vg_{\vk} = (\vk\times \hat{\bf z})/k_f$,
%= (k_y, -k_x, 0)/k_f$,
and a triplet order parameter,
$\Delta_+=-\Delta_- =\Delta$; results for
$\Delta_+\neq -\Delta_-$ and different SO couplings
will be presented elsewhere \cite{abv08a}.

%%%%%%%%%%%%%%%%%%%%%%%%%%%%%%%%%%%%%%%%%%%%%%%%%%%%%%%%%%%%%%%%%%%%%%%%
\begin{figure}[t]
%\centerline{\includegraphics[width=8cm]{boundstateE_supprOP.eps}}
\centerline{\includegraphics[width=8cm]{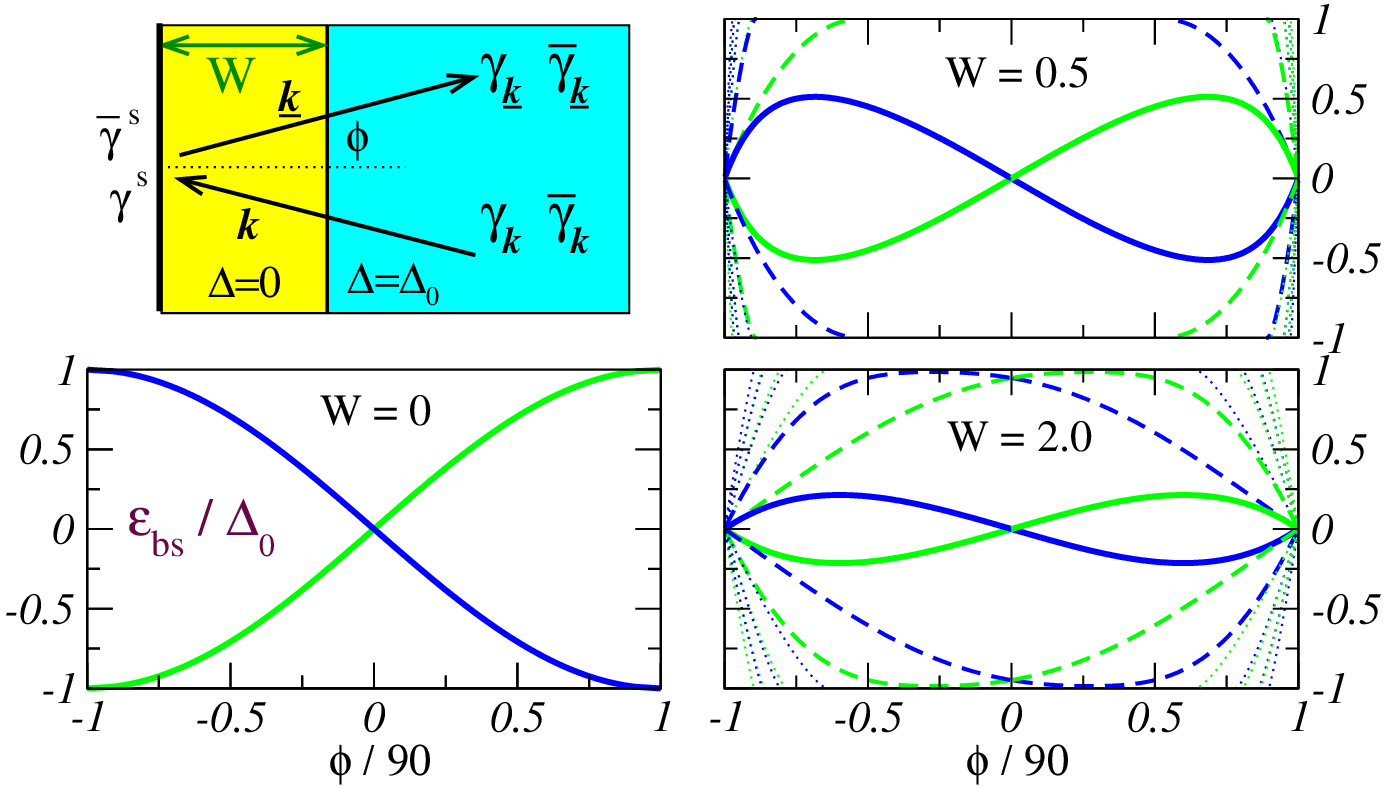}}
\caption{(Color online) Bound state energy as a function of the
impact angle for different widths $W$ (in units of of
$v_f/2\Delta_0$) of the order parameter suppression region. Blue
(green) lines correspond to plus (minus) sign in
Eq.(\ref{eq:ABS-W}). Solid lines: principal mode; broken lines:
higher multiple reflection modes.} \label{fig:2}
\end{figure}
%%%%%%%%%%%%%%%%%%%%%%%%%%%%%%%%%%%%%%%%%%%%%%%%%%%%%%%%%%%%%%%%%%%%%%%%

To obtain insight in the role of the OP suppression, we consider
first a simple model where $\Delta=0$ in a layer of width $W$ next
to the surface, see Fig.~\ref{fig:2}. Trajectories incident at an
angle $\phi$ travel through a non-SC region of an effective width
$2D=2W/\cos\phi$. In this case $\cM = \cot^2 \phi $, the surface
coherence amplitudes gain a phase factor, $\gamma_\pm =
\gamma^\sm{0}_\pm \, e^{i 2\vare D/ v_f}$, $\tilde\gamma_\pm =
\tilde\gamma^\sm{0}_\pm \, e^{i 2\vare D/ v_f}$, and the bound
states are given by $\mbox{Im}^2 (\tilde\gamma^\sm{0}_+ e^{i
2\vare D/ v_f}) = \mbox{Re}^2 (\tilde\gamma^\sm{0}_+ e^{i 2{\vare
D/ v_f}}) \; \cM $,
which yields
    \be \frac{\vare}{\sqrt{\Delta_0^2-\vare^2}} =
    - \tan \left({2 W \, \vare \over v_f \cos\phi} \pm \phi \right).
    \label{eq:ABS-W}
    \ee
Solutions of this equation are shown in Fig.~\ref{fig:2}. The
``principal'' modes with energies away from the continuum edge
contribute the most to the subgap DOS. $W=0$ reproduces the result
of Ref.~\onlinecite{ini07}: each incoming trajectory yields a
bound state at a different energy.
For $W\ne0 $ the main mode
$\varepsilon_{bs}(\phi)$ develops a maximum at
$\varepsilon^\star<\Delta_0$, and we expect a
peak in the surface DOS near $\varepsilon^\star$ due to abundance
of trajectories contributing to $N(\varepsilon^\star)$.

%%%%%%%%%%%%%%%%%%%%%%%%%%%%%%%%%%%%%%%%%%%%%%%%%%%%%%%%%%%%%%%%%%%%%%%%
\begin{figure}[t]
\centerline{\includegraphics[width=8cm]{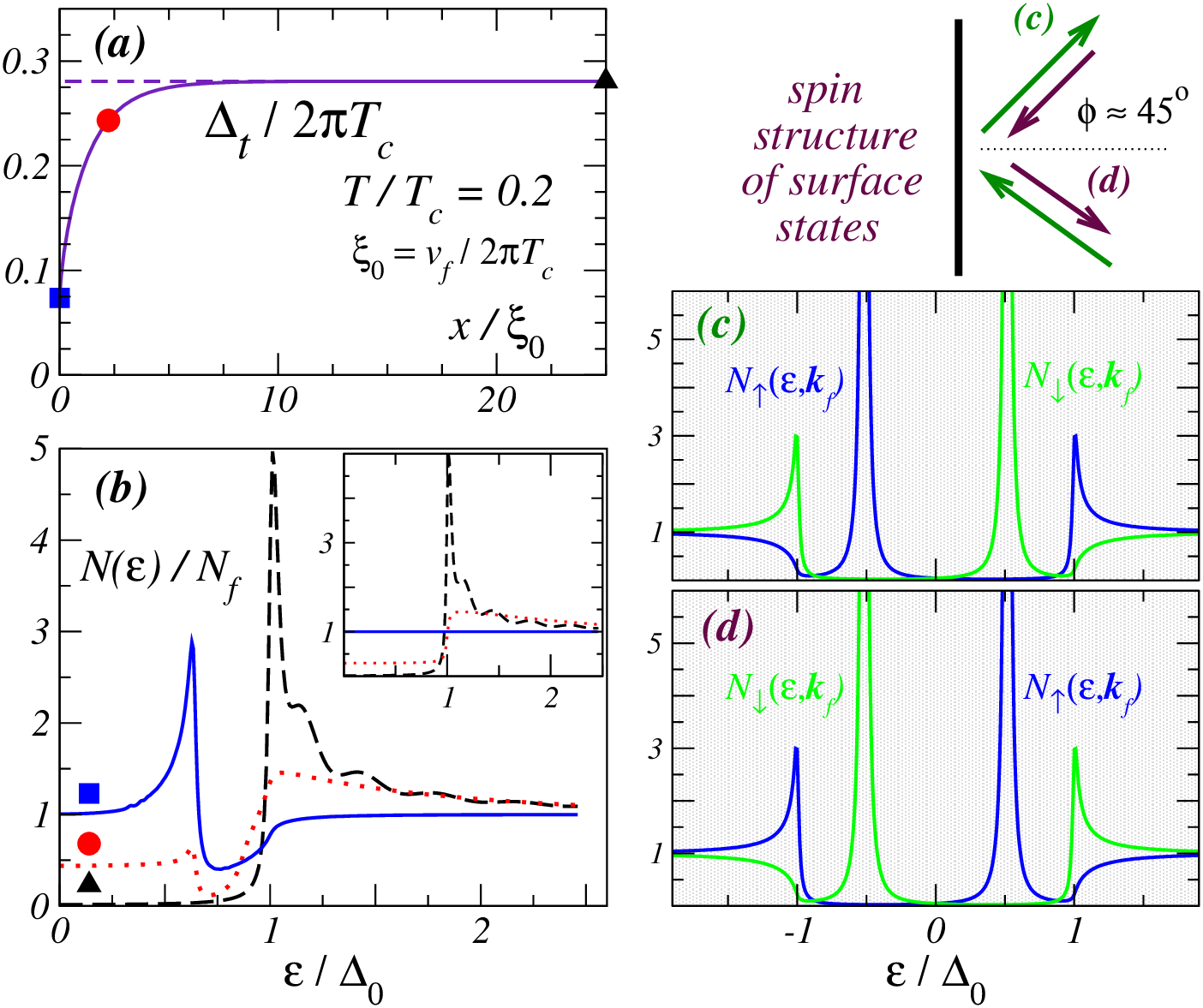}}
\caption{ \label{fig:selfcons} (Color online) Structure of the
surface states for a Rashba triplet superconductor ($\Delta_+=-\Delta_-$).
(a) Order parameter suppression; (b) DOS at
locations indicated in (a); the large sub-gap peaks
are due to the suppression of $\Delta $. Inset: DOS for a
homogeneous order parameter; note the absence of any subgap
features.
(c,d) spin-resolved surface DOS for two trajectories;
$N_{\uparrow}(\epsilon,\vk_f)$ (blue) and
$N_{\downarrow}(\epsilon,\vk_f)$ (green)
correspond to blue and green branches in Fig.~\ref{fig:2}.
}
\end{figure}
%%%%%%%%%%%%%%%%%%%%%%%%%%%%%%%%%%%%%%%%%%%%%%%%%%%%%%%%%%%%%%%%%%%%%%%%
Fully self-consistent solution, shown in Fig.~\ref{fig:selfcons}(b),
confirms this. Note that $\Delta \neq 0$
at the surface, Fig.~\ref{fig:selfcons}(a),
as in other  unconventional superconductors
misaligned with respect to the interface~\cite{buc95a}. Crucially,
self-consistency does yield a peak in the surface DOS below the
gap at a finite energy. Experimentally accessing this peak by
point contact tunneling requires a sufficiently wide tunneling cone
as the feature arises from the trajectories at intermediate
incident angles, see Fig.~\ref{fig:2}.

These ABS
have unusual spin structure. Fig.~\ref{fig:selfcons}(c,d)
shows the spin resolved density of states, $N_{\uparrow\downarrow}
= N \pm N^Z$, where $N$ is the net DOS and $N^\alpha (\epsilon
,\vk_f,\vx )= -\frac{N_f}{2\pi} \mbox{Im} \mbox{Tr}\left\{
    \sigma^\alpha g(\epsilon, \vk_f, \vx)\right\}$. At the interface
    $N^X=N^Y=0$.
The states corresponding to different branches of
Eq.~(\ref{eq:ABS-W}) have opposite spin polarization. Since the
spin polarization changes sign for reversed trajectories, the
Andreev states carry spin current along the interface.

Spin currents exist in
NCS materials because the spin is not conserved, and consequently
precession terms enter the continuity equation,
$\partial_t S^\alpha(\vx)  + \grad\cdot\vPi^\alpha(\vx)
= P^\alpha(\vx)$
\cite{EIRashba:2003}.
Here, the spin density,
$S^\alpha(\vx) =
   \frac{1}{2} \mbox{Tr} \int d\vk \; \sigma^\alpha G(\vk,\vx )$,
the spin current,
$\vPi^\alpha(\vx) =
 \frac{1}{4}\mbox{Tr}\int d\vk \; \{ \sigma^\alpha \, ,\, \vv_{\vk} \}
G(\vk, \vx )$,
and the precession
$P^\alpha(\vx) =
 \frac{1}{2i}\mbox{Tr}\int d\vk \; [\sigma^\alpha \,,\, \vv_{\vk} \cdot \vk]\;
G(\vk ,\vx )$, (where $[\bullet,\bullet]$ is a commutator,
and $\vv_{\vk} =\vk_f/m + \alpha_R(
\hat{\bf z} \times \vsigma )$ is the band velocity), are all given in terms
of
Green's functions
at imaginary relative time $\tau=-i0$.
For the Rashba case, the
precession terms are related to spin currents via the relations
$P^X = -2m\alpha_R \Pi^Z_x $, $P^Y = -2m\alpha_R \Pi^Z_y$, $P^Z =
2m\alpha_R (\Pi^X_x+\Pi^Y_y)$.~\cite{erl05}

%~~~~~~~~~~~~~~~~~~~~~~~~~~~~~~~~~~~~~~~~~~~~~~~~~~~
\begin{figure}
\includegraphics[width=0.85\linewidth]{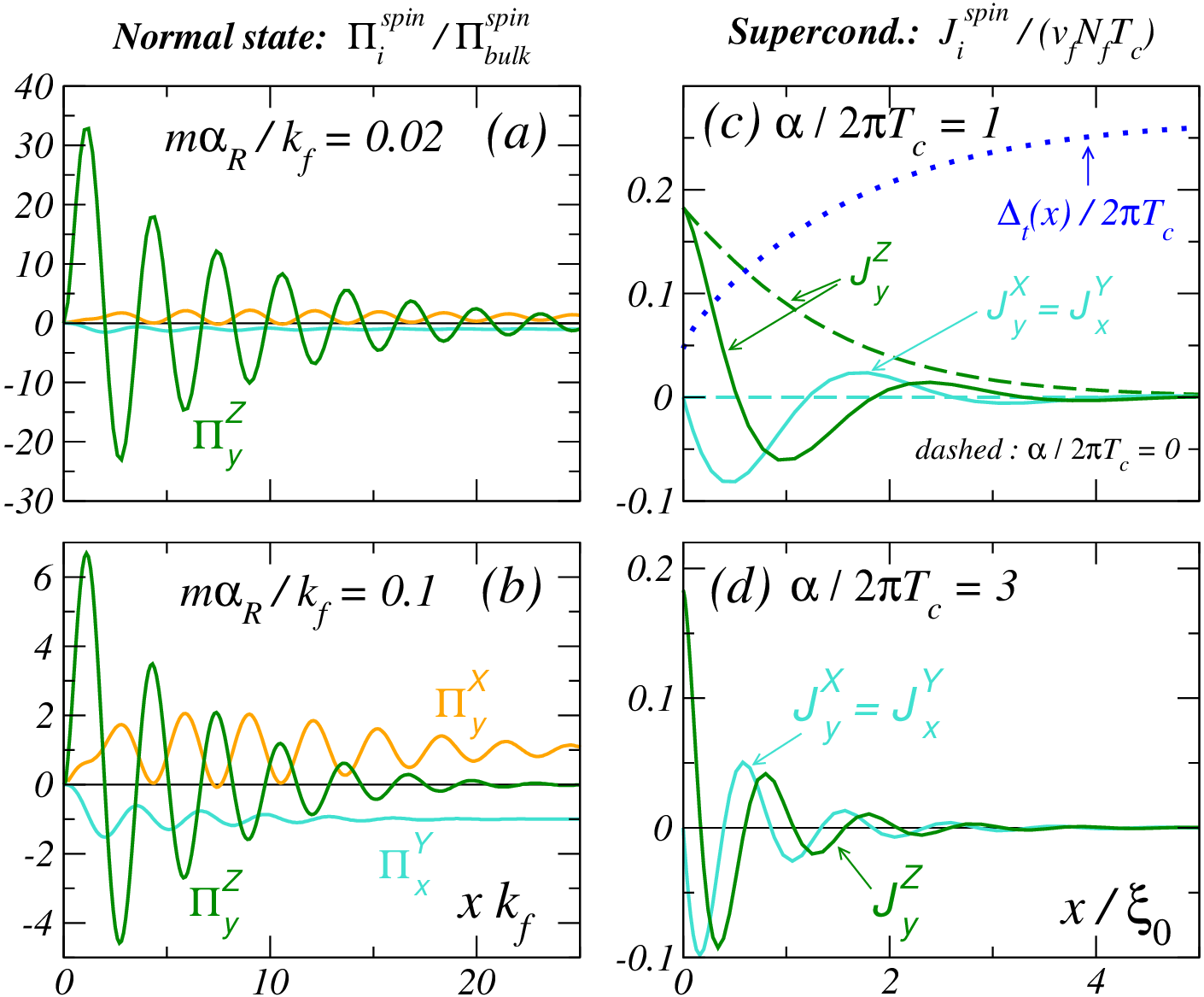}
\caption{\label{fig:Nsc} (Color online) Spin currents near the
surface. Left: Rashba metal. The non-vanishing components in the
bulk are $\Pi^Y_x=-\Pi^X_y$. Near the surface $\Pi^Z_y$ is large,
see text. Right: Spin currents and the order parameter in the SC
state.}
\end{figure}
%~~~~~~~~~~~~~~~~~~~~~~~~~~~~~~~~~~~~~~~~~~~~~~~~~~~

We first consider the spin currents in the normal state. The bulk
value, $\Pi^Y_x = - \Pi^X_y=\Pi^{bulk}_{spin} = m^2\alpha_R^3 /
3\pi$ agrees with Ref.~\onlinecite{EIRashba:2003}. To determine
the surface spin currents we find the Green's function for a
surface modeled as a $\delta$-function barrier at $x=0$ of
strength $U$. The Dyson equation in 2x2 spin space reads $
G^{-1} = [ G^{(0)}]^{-1} - U \delta(x)$, where $[{G}^{(0)}_\vk
]^{-1} = \vare - \xi_\vk -
\alpha_R (\vk \times \hat{\bf z})\vsigma$.
For an impenetrable surface ($U\to\infty$) the
solution is (for fixed $k_y $)\cite{mat95b}
    \be {G}_{k_x k_x'} = {G}^{(0)}_{k_x} 2\pi \delta(k_x-k_x')
- {G}^{(0)}_{k_x} {1\over \int {d p_x\over2\pi}
{G}^{(0)}_{p_x}} {G}^{(0)}_{k_x'}. \label{eq:ngf}
    \ee
We solve Eq.(\ref{eq:ngf}) numerically, and show the normal state
surface spin currents in Fig. \ref{fig:Nsc}(a,b). The most
prominent new feature is a large surface current $\Pi^Z_y$ with
out of plane spin polarization (reminiscent to that in spin Hall bars
\cite{EGMishchenko:2004}) that flows
along the surface, and decays rapidly into the bulk on a Fermi wavelength
scale.
%the scale similar to that of Friedel oscillations. 
This component is related
to $\Pi^Y_x$ via the continuity equation, $\Pi^Z_y(x) =
-1/(2m\alpha_R) \; d\Pi^Y_x(x)/dx$. As a result,  this component
is much greater, by a factor of order $k_f/m\alpha_R$, than the
bulk spin currents in the normal state.

The SC spin current, shown in Fig.~\ref{fig:Nsc}(c,d),
is defined in the quasiclassical method relative to the normal state,
 \bea
\vJ^\alpha \equiv
\vPi^\alpha-\vPi^\alpha_N
=
\int_{-\infty }^{\infty} d\epsilon \;
 n_f(\epsilon)
\big\langle  \vv_f
 N^\alpha(\epsilon, \vk_f, \vx) \big\rangle \,,
    \eea
where $n_f(\epsilon) $ is the Fermi function. The surface-induced
current with out of plane spin polarization is greater than
the normal state current by the factor $\sim T_c E_f^2/\alpha^3$.
The maximal amplitude at the surface is solely determined by the
structure of the SC gap and formally survives even in the limit
$\alpha \to 0$. SC spin currents decay into the bulk on the scale
of the coherence length, much slower than in the normal phase. The
oscillations in Fig. \ref{fig:Nsc}(c,d) are determined by the
spin-orbit strength $\alpha $ and appear due to Faraday-like
rotations of the spin coherence functions along quasiparticle
trajectories.

In summary, we developed a framework for the analysis of
surface bound states and the associated spin currents in
non-centrosymmetric superconductors, and applied it to a system
with a Rashba-type spin-orbit coupling. We found that the
suppression of superconductivity near the surface gives rise
to a finite bias peak in the surface density of states that can
be probed by point contact tunneling. We also showed that large in
amplitude and slowly decaying spin currents with out of plane spin
polarization are carried by these surface states. This opens the
route to future investigations of spin transport in systems
containing superconductors without center of inversion.

\paragraph{Acknowledgements.}
This work was supported by the Louisiana Board of Regents,
and through I2CAM by NSF grant DMR 0645461.

%%%%%%%%%%%%%%%%%%%%%%%%%%%%%%%%%%%%%%%%%%%%%%%%%%%%%%%%%%%%%%%%%%%%%
%%%%%%%%%%%%%%%%%%%%%%%%%%%%%%%%%%%%%%%%%%%%%%%%%%%%%%%%%%%%%%%%%%%%%
%%%%%%%%%%%%%%%%%%%%%%%%%%%%%%%%%%%%%%%%%%%%%%%%%%%%%%%%%%%%%%%%%%%%%
%%%%%%%%%%%%%%%%%%%%%%%%%%%%%%%%%%%%%%%%%%%%%%%%%%%%%%%%%%%%%%%%%%%%%
%\bibliographystyle{apsrev}
%\bibliography{NCS}
%%%%%%%%%%%%%%%%%%%%%%%%%%%%%%%%%%%%%%%%%%%%%%%%%%%%%%%%%%%%%%%%%%%%%
%%%%%%%%%%%%%%%%%%%%%%%%%%%%%%%%%%%%%%%%%%%%%%%%%%%%%%%%%%%%%%%%%%%%%
%%%%%%%%%%%%%%%%%%%%%%%%%%%%%%%%%%%%%%%%%%%%%%%%%%%%%%%%%%%%%%%%%%%%%
%%%%%%%%%%%%%%%%%%%%%%%%%%%%%%%%%%%%%%%%%%%%%%%%%%%%%%%%%%%%%%%%%%%%%

\end{document}